\def\apj{ApJ\ }
\def\apjl{ApJL\ }  
\def\apjs{ApJS\ }
\def\aa{A\&A\ }   
   
\def\etal{{\it et al.~}} 
\def\ha{hereafter~} 
\def\ee{EETDA}
\def\Mpc{$h^{-1}$~Mpc}
\documentstyle[epsf]{l-aa}     

\begin{document}
 
   \thesaurus{12 (12.03.3; 12.12.1) }      
%
\title{ The Supercluster--Void Network. }
\subtitle{ I. The Supercluster Catalogue and Large-Scale 
Distribution}
\author{ Maret Einasto\inst{1}, Erik Tago\inst{1}, Jaak
Jaaniste\inst{1}, Jaan Einasto\inst{1} and Heinz Andernach\inst{2}} 
\institute{ $^{1}$ Tartu Observatory, EE-2444 T\~oravere, Estonia\\
            $^{2}$ IUE Observatory, Villafranca, Apartado 50727, 
                   E-28080 Madrid, Spain} 
\offprints{M. Einasto }
\date{Received ... 1996 / Accepted ... 1996} 
\maketitle
\markboth{Einasto \etal: The Supercluster--Void Network I}{}
\begin{abstract} 
We investigate the distribution of superclusters and voids using a
new catalogue of superclusters of rich clusters of galaxies which
extends up to a redshift of $z = 0.12$.  The new catalogue contains
220 superclusters of rich clusters, of which 90 superclusters
have been determined for the first time. Among them there are
several very rich superclusters, containing at least eight member
clusters.

We demonstrate that two thirds of very rich superclusters are
concentrated to a Dominant Supercluster Plane which is situated at
a right angle with respect to the plane of the Local Supercluster and
adjacent nearby superclusters.

We apply several methods to estimate the characteristic distance
between superclusters. The results indicate consistently the
presence of a quite regular supercluster-void network with scale of
{$\approx 120$} \Mpc.

Comparison with random supercluster catalogues shows significant
differences between spatial distributions of real and random
superclusters. 

We determine the selection function of the sample of clusters and suggest
that the mean true space density of Abell clusters is 
 $2.6 \times 10^{-5} h^3 {\rm Mpc}^{-3}$,
twice the conventionally used value
\footnote{
Table A2 is only available in electronic form
at the Centre des Donn\'ees Strasbourg 
via anonymous ftp to cdsarc.u-strasbg.fr (130.79.128.5)
or via http://cdsweb.u-strasbg.fr/Abstract.html}
.

\keywords{ cosmology: observations 
--- large-scale structure of the Universe }   
\end{abstract}

\maketitle

\section {Introduction}

Galaxies and systems of galaxies form due to initial density perturbations
of different scale. Short perturbations with a 
wavelength of several Mpc
give rise to the formation of individual galaxies and small systems of
galaxies, medium scale perturbations lead to the formation of
clusters of galaxies, and so on. Perturbations of a characteristic
scale of $\sim 100$\ \Mpc\ can be related to superclusters of
galaxies. Still larger perturbations have much lower amplitude and
thus they only modulate densities and masses of smaller systems
(Frisch \etal 1995). Therefore superclusters of galaxies are the
largest relatively isolated density enhancements in the Universe.

The presence of superclusters is known since the pioneering studies of Shapley
(1930). The nearest example is the Local Supercluster with the Virgo cluster
as the central cluster (de Vaucouleurs 1956). Other nearby examples are the
Perseus-Pisces supercluster which consists of the Perseus chain of rich
clusters, and the Coma supercluster with the Coma cluster and A1367 forming
its double center. The distribution of galaxies in superclusters is
filamentary, these filaments can contain as density enhancements groups 
and clusters galaxies of different richness (Gregory and Thompson 
1978, J\~oeveer \etal 1978, Einasto \etal 1984).

Superclusters are not completely isolated in space. Galaxy and cluster
filaments connect neighbouring superclusters to a single network.  Filaments
joining the Local and Perseus-Pisces superclusters were noticed by Einasto
\etal 1980, and filaments joining the Local and Coma superclusters by
Zeldovich, Einasto and Shandarin (1982) and Tago, Einasto and Saar (1984,
1986).  A section of the Great Wall (Geller and Huchra 1989) is a filamentary
system which joins the Coma and Hercules superclusters (Lindner \etal 1995).

We shall use the term supercluster-void network for the web of
filaments, clusters, and voids which extends over the whole
observable part of the Universe.  The formation of a filamentary
web of galaxies and clusters is predicted in any physically
motivated scenario of structure formation (of recent works we
mention studies by Bond, Kofman \& Pogosyan 1996 and Katz \etal
1996).  Properties of this network depend on the density
perturbations of medium and large wavelengths. Thus the study of
the properties of the supercluster-void network yields information
on the shape of the initial power spectrum on these wavelengths.
Of particular interest is the region of transition from the
Harrison-Zeldovich spectrum with positive power index $n=1$ on very
large scales to galactic scales with negative effective power index
$n\approx -1.5$. In this wavelength region differences between
various structure formation scenarios are the largest.

Our present series of papers is devoted to the study of the
properties of the supercluster-void network.  Superclusters can be
determined using the appropriately smoothed density field, or using
discrete tracers, such as galaxies or clusters of galaxies, and
applying the clustering analysis. In both cases superclusters can
be defined as the largest non-percolating systems of galaxies or
clusters of galaxies.  Decreasing the threshold density or
increasing the neighbourhood radius we get already a percolating
system -- the supercluster-void network.  Both galaxies and
clusters are concentrated to superclusters and trace similar
high-density regions of the Universe (Oort 1983; Bahcall 1991).  In
detail the distributions are different, since clusters of galaxies
trace only compact high-density regions -- the skeleton of the
structure. The use of galaxies as tracers of superclusters is
limited to relatively small distances as catalogues of redshifts of
galaxies which cover a large fraction of the sky are not deep and
complete enough yet. On the contrary, the catalogues of rich
clusters of galaxies by Abell (1958) and Abell \etal (1989, ACO),
which cover the whole sky out of the Milky Way zone of avoidance,
are thought to be fairly complete up to distances of several
hundred megaparsecs. Thus most supercluster studies were based on
these catalogues of clusters of galaxies.

Catalogues of superclusters using  clusters as structure tracers
have been compiled by Bahcall and Soneira (1984); Batuski and Burns
(1985); West (1989); Postman, Huchra and Geller (1992). The first
whole-sky supercluster catalogues were prepared by Zucca \etal
(1993, hereafter ZZSV); Einasto \etal (1994, \ha EETDA); and
Kalinkov and Kuneva (1995), the last one uses mainly clusters with
estimated redshifts.

In the study of the distribution of superclusters it is of central
importance to know whether it deviates from a random distribution, 
and if yes, 
whether the supercluster distribution defines a certain scale in
the Universe.  These questions were addressed already by Oort
(1983).  Subsequent studies have shown the presence of some
regularities in the distribution of superclusters.  Zeldovich,
Einasto and Shandarin (1982) and Tully (1986, 1987) demonstrated
that nearby superclusters are concentrated to a plane which almost
coincides with the plane of the Local Supercluster. This
concentration of superclusters forms a wall dividing two huge
voids, the Northern and Southern Local voids (Einasto and Miller
1983; Lindner \etal 1995). Tully \etal (1992) showed that several
superclusters are almost perpendicular to Local Supercluster plane.
EETDA suggested that superclusters and voids form a quite regular
network with a characteristic distance between superclusters of
about $110 - 140$ \Mpc. A similar scale was found by Mo \etal
(1992) in the distribution of clusters of galaxies; the value is
also close to that found by Broadhurst \etal (1990) for the
distance between peaks in the redshift distribution of galaxies in
a pencil-beam survey of galaxies and. The scale of about 100 \Mpc~
has    also been found in the distribution of QSO absorption line
systems (Quashnock, Vanden Berk and York 1996). These results
suggest the presence of a peak in the power spectrum of density
fluctuations at the corresponding wavelength (Einasto and Gramann
1993; Frisch \etal 1995).  An excess power in the power spectrum of
galaxies of the Las Campanas redshift survey has been detected at
this scale by Landy \etal (1996).

In recent years the number of clusters with measured and
re-measured redshifts has been increased considerably. Thus a new
and more detailed analysis of the distribution of clusters and
superclusters is possible. In this series of papers we shall
construct a new catalogue of superclusters,  study the
large-scale distribution of superclusters (the present paper),
determine the correlation function and the power spectrum of
clusters of galaxies (Einasto \etal 1996a,b,c), investigate the
form and orientation of superclusters (Jaaniste \etal 1996),
compare the distribution of clusters and superclusters of galaxies
with the distribution of similar objects in numerical simulations
(Frisch \etal 1996), and investigate consequences of these results
to scenarios of structure formation (Einasto \etal 1996a,b).

The present paper is arranged as follows.  Section 2 presents a new
catalogue of superclusters up to $z = 0.12$.  Redshift data are
available for 2/3 of the clusters within this distance limit. We
use this limit in order to include very rich superclusters missed
in the earlier version of the catalogue (EETDA).   In addition, we
apply improved distance estimates for clusters without observed
redshifts (Peacock and West 1992). In Section 3 we determine the
selection function and mean space density of clusters. In Section 4
we describe catalogues of randomly located superclusters.  In
Section 5 we use the catalogue of superclusters to describe and
analyse the structures delineated by superclusters on large scales,
and compare the spatial distribution of rich and poor superclusters
and isolated clusters.  In Section 6 we analyse the sizes of
voids defined by rich clusters from systems of various richness. In
Section 7 we calculate the characteristic distance between the
largest systems. In Section 8 we study the distribution of
superclusters of different richness in void walls. Section 9 gives
a summary of principal results.

We denote with $h$ the Hubble constant in units of 100 km s$^{-1}$
Mpc$^{-1}$.

\section {The catalogue of superclusters} 

\subsection{The cluster data} 

The Abell--ACO catalogue includes 2712 northern clusters originally
published by Abell (1958), 1364 rich southern clusters that are
counterparts to the Abell clusters and 1174 supplementary poor
southern clusters (Abell, Corwin and Olowin 1989).  Some rich
clusters are duplications, therefore the combined Abell--ACO
catalogue includes at most 4069 rich clusters. In this paper we use
only these rich clusters of the Abell--ACO catalogue and call them
simply as clusters.

We are updating redshift data for Abell--ACO clusters continuously
using all available sources including some unpublished redshifts.
The present discussion reflects our dataset as of May 1995.  A
catalogue of published redshifts and velocity dispersions for
Abell--ACO clusters, including supplementary clusters, is in
preparation (cf. Andernach, Tago and Stengler-Larrea 1995). For
clusters without observed redshift a photometric estimate of the
distance is given using the correlation between redshifts and
magnitudes of cluster galaxies (Peacock and West 1992). 
The errors of estimated redshifts are about 27\% for the northern
(Abell) and 18\% for the southern (ACO) clusters which are
considerably higher than errors for spectroscopically measured
redshifts. The redshifts have been corrected to the rest frame of
the Local Group ($\Delta z = 0.001 \sin l \cos b$) and for the
expansion effects.  The expansion correction depends on the adopted
model and density parameter of the universe. We have used a
correction which corresponds to a closed universe
($\Omega_0+\Omega_{\Lambda}=1$) and a value of the density
parameter, $\Omega_0\approx 0.7$.  Results depend on the particular
value of the density parameter only very weakly.

For a number of clusters published redshifts obviously belong to a
foreground or background galaxy (some of them are marked by ACO and
Struble and Rood 1991, and also by Dalton \etal 1994).  We have used
estimated redshifts instead of poorly observed ones if
$|\log(z_{obs}/z_{est})| > 0.3$ and if the number of measured
galaxy redshifts per clusters was $n_z < 3$.  The influence of such
clusters on our catalogue will be discussed later.

To compile the supercluster catalogue we extracted from the whole
Abell--ACO catalogue a spatially limited sample up to a distance $z
= 0.12$. This sample contains 1304 clusters, and includes clusters
of all richness classes. Of these clusters 2/3 have measured
redshifts. We have included in our study clusters of richness class
0. Arguments for this were already discussed by \ee.  Possible
projection effects discussed by Sutherland (1988), Dekel \etal
(1989) and others are not crucial for the present study as we are
mostly interested in the distribution of clusters on much larger
scales (cf. \ee).

\subsection{ Supercluster finding procedure}

Superclusters have been determined by the clustering (or
friends-of-friends) algorithm (Huchra and Geller 1982; Press and
Davis 1982; Zeldovich, Einasto and Shandarin 1982).  Clusters are
searched for neighbours at a fixed neighbourhood radius; objects
having distances between each other less than this radius are
collected to a system.  We use the same neighbourhood radius as in
\ee, 24~ \Mpc. \ee\ showed that at neighbourhood radii up to
about 16 \Mpc\ the cores of individual superclusters start to form;
at radii larger than 30 \Mpc\ superclusters begin to join into
percolating agglomerates.  At the radius of about 24~ \Mpc\
superclusters are the largest still relatively isolated density
enhancements in the Universe. Our analysis shows that the main
results do not change if we use the neighbourhood radius in the
interval of 20 -- 28~ \Mpc.

In some cases the clustering radius used here is too large, and
forces clusters to join into large aggregates which probably cannot
be considered as single superclusters. One example for this is the
Shapley supercluster that will be discussed by Jaaniste \etal (1996).

\subsection{ The catalogue of superclusters}

We include in the catalogue of superclusters all systems with at
least two member clusters. We shall use the term {\it multiplicity}
$k$ for the number of member clusters in a supercluster. The
distance limit is set at $z = 0.12$; in this volume there are in
total 220 superclusters (for the neighbourhood radius 24~ \Mpc).
The distribution of multiplicities of the superclusters in our
catalogue is shown in Figure~1. Here we plot also isolated
clusters.  Complete data on superclusters having at least four
members (multiplicity, centre coordinates, list of member clusters
and identifications with previous catalogues) are given in Table~A1
in the Appendix, the whole catalogue is presented in electronic
form in Table A2.  Clusters for which only estimated redshifts are
available are appended by a letter $e$.

A number of superclusters have  well-known previous
identifications.  These are given in column (7) of Table~A1. Their
designations are usually based on the constellation on which the
supercluster members are projected. In the case of rich,
well-determined superclusters without previous identifications we
assigned new identifications using the same system. If there were
more than one supercluster projected on the same constellation, we
added the letters A, B, and so on (in order of increasing $z$).
Otherwise, if the supercluster members were projected on more than
one constellation, we used a double name.

About 1/3 of the clusters in our sample have estimated redshifts
only (437 of 1304 clusters).  The median distance of clusters with
measured redshifts (230~ \Mpc) is smaller than that of clusters with
estimated redshifts (300~ \Mpc), which reflects the better
completeness in redshift measurements for nearer clusters.

In order to see the influence of the use of clusters with estimated
redshifts on our catalogue, we performed the cluster analysis
using only clusters with measured redshifts. We searched for
systems using the same neighbourhood radius as before, 24~ \Mpc. As
a result we obtained a test catalogue of superclusters with 136
systems. All the superclusters containing less than two members
with measured redshifts disappeared, of course, after this
procedure.  However, the remaining superclusters appeared to be
surprisingly stable: almost all systems with at least two clusters
with measured redshifts were found also in this test
catalogue, and only a few clusters with measured redshifts were
excluded from systems.  One supercluster, the Aquarius
supercluster (SCL 205), was split up into two subsystems.

Thus we consider all the superclusters with less than two members
with measured redshifts as supercluster candidates. These
superclusters have a letter $c$ to its catalogue number. We
also marked those clusters with measured redshifts that were
eliminated from systems determined by clusters with measured
redshifts only, as described above.

\begin{figure}[htbp] 
\epsfysize=8cm
\hbox{\epsfbox[115 115 360 460]{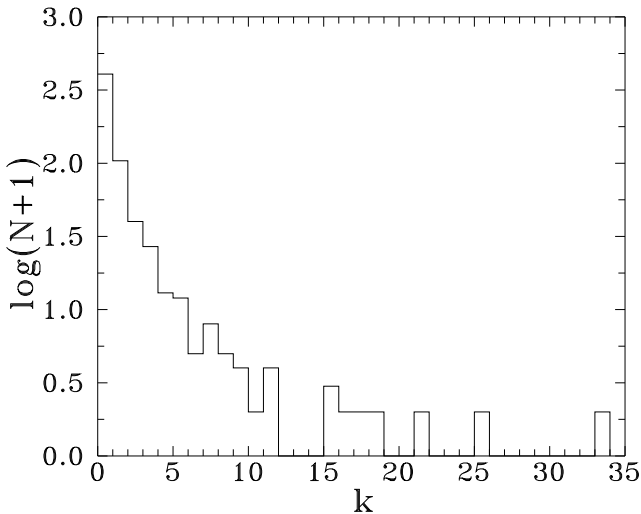}}
\caption { The distribution of supercluster multiplicities 
 for the neighbourhood radius $R = 24 $ \Mpc. Isolated clusters 
 ($k = 1$) are included for comparison.} 

\end{figure}

Of the 220 systems in the new catalogue, 50 superclusters are
identical with superclusters in the previous catalogue, 80 have changed
the multiplicity (in most cases these superclusters have gained or
lost 1 - 2 members due to newly measured redshifts).
The catalogue contains 25 previously unreported superclusters within the
distance of $d < 300$ \Mpc;  all 65 superclusters beyond 300
\Mpc\ are reported here for the first time.
As seen from these numbers, our regular updating of the catalog has
lead to a considerable improvement.  In addition, our analysis showed
that the large scale structures delineated by superclusters from the
present and previous catalogues are almost identical in the nearby
volume covered by both catalogues.

\begin{figure}[htbp] 
\epsfysize=8cm
\hbox{\epsfbox[75 75 360 460]{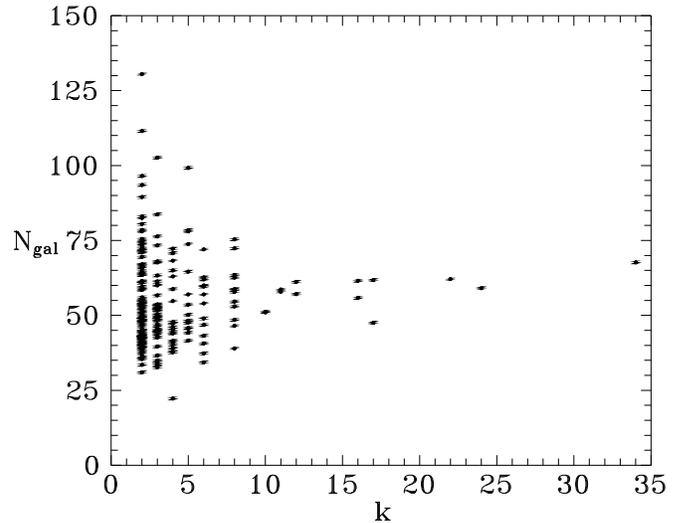}}
\caption {Mean number of galaxies in clusters belonging to 
superclusters of multiplicity $k$. }
\end{figure}

We divide superclusters into several richness classes.  We call
superclusters with less than 4 members as {\it poor}, and those
with 4 or more members as {\it rich}. Rich superclusters are
divided into subclasses: superclusters with 4 - 7 members are
called as {\it medium rich}, and those with 8 or more members as
{\it very rich}.  About half of the 220 superclusters of the
catalogue are cluster pairs; the catalogue contains 53 medium rich
superclusters, and 25 very rich superclusters. Very rich
superclusters represent the regions of the highest density in the
Universe.  They contain 25\% of all clusters and over 30\% of all
supercluster members. Of these very rich superclusters 4 have been
catalogued for the first time.  These are the Draco (SCL 114, $k =
16$), the Caelum (SCL 59$_c$, $k = 11$), the Bootes A (SCL 150, $k
= 10$), and the Leo -- Virgo (SCL 107, $k = 8$) superclusters.  In
the following Sections we shall compare the spatial distribution of
superclusters of different richness.

\begin{figure}[htbp] 
\epsfysize=14.cm
\hbox{\epsfbox[130 90 440 540]{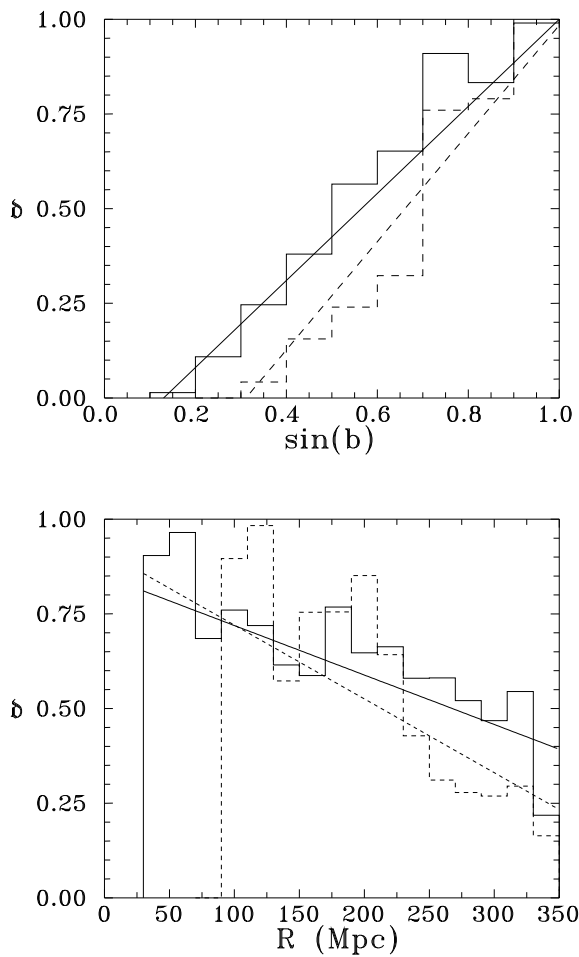}}
\caption {Selection functions for clusters. In the upper panel the
 density of clusters is shown as a function of the galactic
latitude, $\sin b$, in the lower panel as a function of distance from
the observer, $r$. Solid histograms are for all clusters,
dashed histograms for     clusters from very rich superclusters. Straight lines
show the linear approximation of the selection function. The curves
are normalised to 1 for the galactic poles and for zero distance to
the observer. }
\end{figure}

Supercluster masses are evidently larger when they contain more
galaxies.  To check the relationship between the supercluster
richness and the number of galaxies contained in a supercluster we
plot in Figure~2  the mean number of galaxies in
superclusters 
against supercluster multiplicity. We used the Abell count of
galaxies ($C$ in ACO) as the number of galaxies per cluster.
Clearly, the mean number of galaxies in clusters located in
superclusters of different multiplicity is practically constant.
This test shows that the supercluster multiplicity is
an indicator of the mass of the supercluster (see also Frisch \etal
1995).
An example supported by actual observations is the
Shapley supercluster, the richest supercluster in our catalogue.
It contains the richest clusters in the
volume under study and  a large number of X-ray emitting clusters
which indicate the presence of a deep potential well in this
supercluster (Breen \etal 1994; \ee).

\begin{figure*}[htbp]
\epsfysize=14.cm
\hbox{\epsfbox[70 60 350 450]{f4.ps}}
\caption {The distribution of clusters in supergalactic
coordinates in slices of thickness $d = 100$ \Mpc\  in the
supergalactic $X$ direction. 
Clusters belonging to very rich superclusters are
denoted with filled circles; clusters, belonging to medium rich
superclusters -- with empty circles, and
isolated clusters and members of poor superclusters
are plotted with small dots. The first and last slices are thicker
since due to the use of the spherical volume outlying slices contain
less clusters.}
\end{figure*}

\subsection{ Notes on very rich superclusters}

First we give some notes on previously known superclusters.

The {\it  Shapley supercluster} (SCL 124), first described by 
Shapley in (1930), is certainly the most prominent supercluster in
the region under study (ZZSV). This supercluster contains the
richest Abell clusters in the area studied,  and a number of $X$-ray
clusters (Quintana \etal 1995 and references therein).  
This supercluster is located approximately 140~ \Mpc\  from
us, bordering the farther side of the Northern Local void (\ee,
Lindner \etal 1995). 

The {\it Virgo--Coma supercluster} (SCL 111) with 16 members forms a
wall between two voids. Of these 16 clusters 6 have estimated
redshifts about 1.5 times larger than are their (poorly)
observed redshifts. Thus a possible alternative interpretation of the
data is that some of the clusters are more distant, and the
measured redshifts belong to foreground galaxies in the region of
this supercluster. If we discard these clusters then the
supercluster contains at least 8 members and still meets our
criterion for very rich superclusters.

The {\it Horologium--Reticulum supercluster} (SCL 48), the longest
and the second richest supercluster in the previous catalogue,
has been split into subsystems containing now 26 members
instead of 32 (\ee) (see Table~1), being still the second most rich
supercluster in the new catalogue but not the longest one (Jaaniste
\etal 1996).

Now we comment on those very rich superclusters ($k\geq~8$) in our 
catalogue which were not previously reported.

The {\it Draco supercluster} \ (SCL 114) has 16 members, all with
measured redshifts, being one of the richest superclusters in the
region under study.  The Draco supercluster lies at a distance of
300~\Mpc\ on a side of a void of diameter of about 130~\Mpc, the
near side of which  is determined by the Ursa Majoris supercluster.
The Draco supercluster is one of the most isolated very rich
superclusters in our catalog. However, being located 
near the distance limit of
our sample this supercluster might have a neighbour farther
away. The shape of this supercluster resembles a pancake with axis
ratios 1:4:5 (Jaaniste \etal 1996).

The {\it Bootes A supercluster}\ (SCL 150) borders a giant void on
the farther side of the Bootes supercluster which separates this
void from the Bootes void. Nine  of the ten members of this
supercluster have measured redshifts.
                          
The {\it Leo--Virgo supercluster} (SCL 107) has 8 members, six of
them have measured redshifts. This supercluster borders the same
void as SCL 111.

The {\it Caelum supercluster candidate} (SCL 59c) 
borders the same void as the Fornax--Eridanus supercluster and is
seen in Figure~3 by Tully \etal (1992) as a density enhancement.
However, a word of caution is needed: only two of the 11 members of
this supercluster have measured redshifts.

The {\it Fornax--Eridanus supercluster candidate} (SCL 53c) 
too consists mostly of clusters with estimated redshifts. The
multiplicity of this supercluster may  change when new
redshift data for rich clusters in this region become available.

\begin{figure*}[htbp]
\epsfxsize=14cm
\hbox{\epsfbox[50 50 450 550]{f5.ps}}
\caption {The distribution of clusters belonging to very rich
superclusters in supergalactic coordinates. Supercluster
identifications are given. 
}
\end{figure*}

\section{Selection functions and the mean volume density of clusters}

In this Section we study the influence of selection effects 
on the distribution of clusters and superclusters,
and on the space density of clusters. 

The probability to detect a cluster at a certain location depends
on the galactic obscuration and on the distance of the cluster. To
investigate the selection effects we determined the volume density
of clusters of galaxies in bins of spherical shells of thickness of
$20$ \Mpc\ and
in bins of  {$\Delta\sin b=0.1$} ($b$ is the galactic latitude).  Results are
shown in Figure~3, separately for all clusters and for the
population of clusters in very rich superclusters with at least 8
members. The distributions are given for all clusters, but the selection
effects are similar for only the clusters with measured redshifts
(Einasto \etal 1996b).

This Figure shows that the dependence of the space density of
clusters on distance and on $\sin b$ is almost linear. Thus we can
represent the selection effects by linear laws: $D(r)=
d_{0}-d_{1}(r/r_{1})$, and $D(\sin b)=s_{0} + s_{1}\sin b$, where
$d_{0}$, $d_{1}$, $s_{0}$, and $s_{1}$ are constants, and $r_{1}$
is the limiting radius of the sample.  Both for {\it all} clusters
and for those in {\it very rich} superclusters, corrected for
incompleteness and Galactic extinction, we find: $d_{0}=1$,
$d_{1}=0.5$. The latitude dependence is given by the value
$s_{0}=\sin b_{0}$, at which the density is equal to zero. For
samples of all clusters and clusters in very rich superclusters we
get $s_{0}=0.12$, and $s_{0}=0.38$, respectively.

These data were used also to derive the mean number density of
clusters in space. In order to be left with 1304 clusters in the
volume under investigation and with the above selection function,
we actually need approximately 9000 clusters in a cube of side
700~\Mpc.  Thus the mean density of Abell--ACO clusters in space,
corrected for incompleteness and Galactic extinction, is 26 per
cube of side-length 100~ \Mpc, or $2.6 \times 10^{-5} h^3 {\rm
Mpc}^{-3}$, or approximately twice the estimate by Bahcall \& Cen
(1993). This estimate of the space density of clusters is
consistent with the results by Postman \etal (1996) obtained from
the study of distant clusters.

This calculation shows that selection effects are important in
deriving the density of clusters in space.  The comparison with
random supercluster catalogues also shows that in low galactic
latitudes the multiplicity of superclusters is distorted as some
supercluster members are not visible. This explains the observed
fact that the density of rich and very rich superclusters decreases
toward galactic equator more rapidly than the density of poor
superclusters, see Figure~3.

\section{ Random supercluster catalogues}

In order to compare the spatial distribution of clusters and
superclusters with random distributions and to investigate the
influence of the selection effects on the number density of
clusters we generated two sets of randomly located superclusters.

In both sets the number of clusters was the same as in
the observed catalogue (approximately 1300), they occupy the same
volume, and are combined into superclusters that have multiplicity
distribution similar to that of real
superclusters.  First we generated supercluster centres, and then
supercluster members around each centre. The radius of
superclusters was chosen in accordance with observations: 10~\Mpc\ 
and 20~\Mpc, for poor and rich superclusters, respectively (\ee,
Jaaniste \etal 1996).  Clusters outside the sphere of radius,
$r>r_{1}$, and near the Galactic plane,  {$|\sin b| < \sin
b_{0}$}, were excluded. The selection effects were taken into
account in two different ways.

In the first set of random catalogues centres of superclusters and
locations of isolated clusters were generated using a censored
random distribution: in order to avoid overlapping of the
superclusters a minimum distance of 40~ \Mpc\ was chosen between
these centres and a minimum distance of 24~ \Mpc\ was chosen between
clusters that did not belong to the superclusters, as in the case
of real isolated clusters. This set of random supercluster
catalogues was generated without taking into account detailed
selection effects (i.e. clusters were absent from the zone of
avoidance but the changes in the mean density of clusters in
distance and in galactic latitude were ignored).

In the second set of random catalogues we took into account the
selection functions derived from the observed sample of all
clusters.  The locations of the supercluster centres and isolated
clusters were generated completely randomly,  the number of
members for each supercluster was generated according to both the
multiplicity function of real superclusters, and to the selection
functions which determined the probability to find a cluster at a
given galactic latitude and distance from the observer.  The number
of clusters that were generated but not included to the catalogue
due to selection effects gives us an estimate of the real number of
clusters in the volume under study (see the previous Section).
The final multiplicity of superclusters was determined using the
clustering algorithm and a neighbourhood radius of 24~\Mpc.

To check the validity of the selection function procedures we
calculated the density distribution of clusters for both sets of
random supercluster catalogues.  As expected, the dependence of the
cluster density on the galactic latitude and distance was similar
to the observed one in the case of the second set of random
catalogues, while in the case of the first set this dependence was
much weaker.  The multiplicity distribution of superclusters for
the second set of models was almost identical to the multiplicity
distribution in the real catalogue.  We shall discuss the influence
of differences in the selection function on the tests applied in
the present paper in corresponding Sections.

\section{Distribution of  superclusters}

In this Section we study the overall distribution of 
superclusters.  In Figures 4 and 5 we show the distribution of
clusters in supergalactic coordinates. In Figure~4 all clusters are
plotted in slices of 100 \Mpc\ thickness.  In Figure~5 we plot only
clusters belonging to very rich superclusters, in the lower panels
of this Figure clusters from the Southern and Northern sky are
given separately.  Clusters with estimated redshifts (members of
the supercluster candidates) are also included.

Figures 4 and 5 show that the network of superclusters and voids extends
over the entire volume displayed. Superclusters are
separated by huge voids.  For example, the Hercules (SCL 160) and the Shapley
(SCL 124) superclusters border the Northern Local void; the Hercules, the
Bootes (SCL 138) and the Corona Borealis (SCL 158) superclusters surround the
Bootes void, that is bordered by the Draco supercluster (SCL 114) in its far
side. In the Southern sky the Sculptor supercluster (SCL 9) forms the farther
wall of the Sculptor void, to name only the most well--known voids.

The distribution of X-ray emitting clusters from the ROSAT survey 
(Romer \etal 1994) shows essentially the same structures.  
The excess of ROSAT clusters in the region of the Pisces--Cetus
supercluster and in the Sculptor wall are seen particularly well.

\subsection{Supercluster sheets and chains}

Figures 4 and 5 suggest that superclusters are not distributed
homogeneously. Most of very rich superclusters are located 
along rods of a quasi-regular rectangular cubic lattice with almost
constant step, and form elongated structures -- chains.  These
chains are almost parallel to axes of supergalactic coordinates.
The whole distribution of clusters along rods is essentially
one-dimensional.  One possibility to give a quantitative
description of the supercluster chains is to use the fractal
dimension, $D = 3 - \gamma$, where $\gamma$
is the slope of the correlation function expressed in log--log form
(Coleman and Pietronero 1992).  On small scales the slope of the
cluster-cluster correlation function characterises the fractal
dimension of superclusters themselves, on larger scales up to about
90 \Mpc\ the slope is determined by the shape of supercluster
systems. On these scales the fractal dimension determined for all
clusters is $D_{all\ cl} \approx 2$.  This value coincides well
with the correlation fractal dimension for galaxies on large scales
outside clusters (Einasto 1991, Di Nella \etal 1996). The
correlation fractal dimension calculated for clusters that belong
to very rich superclusters is smaller, $D_{scl8}
\approx 1.4$. Thus structures delineated by very rich superclusters
are more one-dimensional than two-dimensional as in the case of
structures defined by all clusters.

Several data sets suggest that giant structures seen in the
Southern and Northern sky may be connected, and superclusters form
sheets or planes in supergalactic coordinates.  One example of such
connection is the Supergalactic  {Plane}, which contains the Local
Supercluster, the Coma Supercluster, the Pisces--Cetus and the
Shapley superclusters (Einasto and Miller 1983; Tully 1986 and 1987;
Tully
\etal 1992, \ee). This aggregate separates two giant voids -- the
Northern and the Southern  {Local} supervoids (\ee).

\begin{figure}[htbp] 
\epsfysize=19.cm
\hbox{\epsfbox[150 150 540 640]{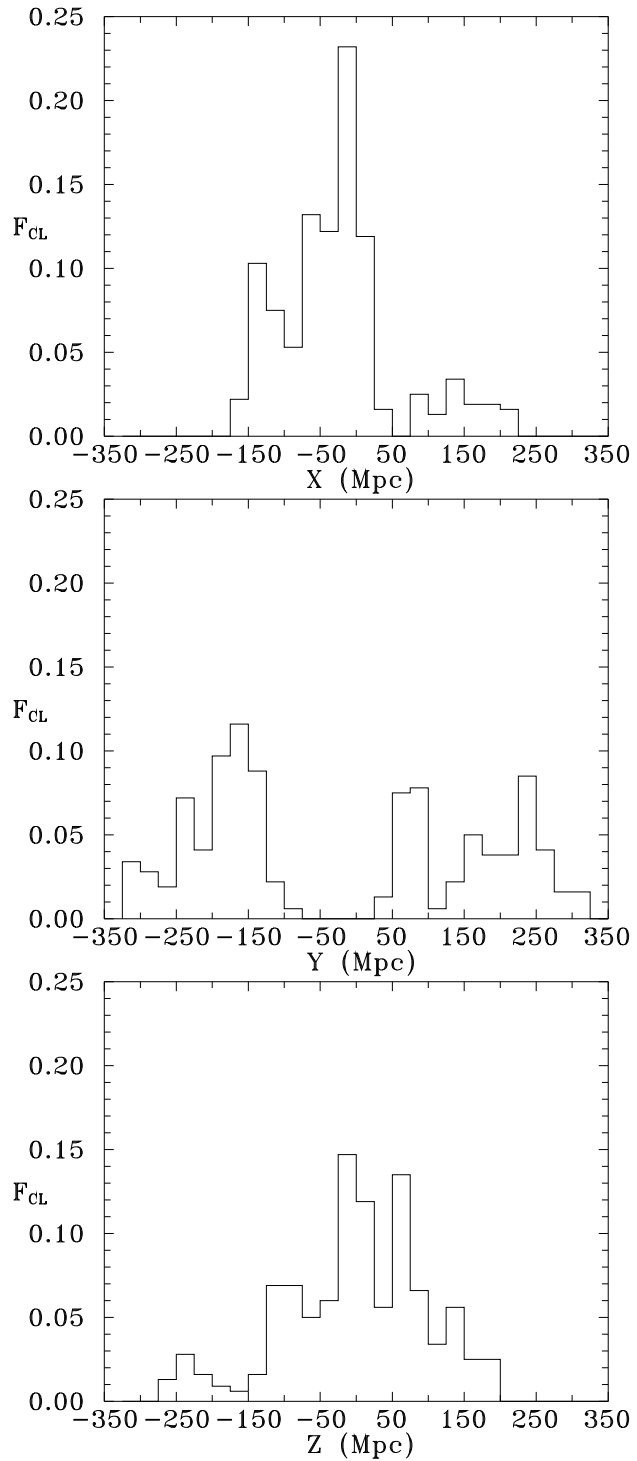}}
\caption {The distribution of member clusters 
of very rich superclusters along supergalactic
coordinates $X$, $Y$, and $Z$. }
\end{figure}

The search of galaxies in the zone of avoidance has provided further
evidence that some other very rich superclusters may be connected
through the zone of avoidance.  Kraan-Korteweg, Fairall and
Balkowski (1995) found that there may be a chain of galaxies in the
zone of avoidance forming a bridge between the Shapley
concentration and the Horologium--Reticulum supercluster.  This
bridge, if real, borders the Southern Local supervoid and connects
chains of superclusters parallel to the Supergalactic plane.

\vskip 5mm

\subsection{The Dominant Supercluster Plane}

The visual impression from Figure~5 is that the upper right panel
(sheet $-75 \leq X \leq 25$~\Mpc) contains most of the members of rich
superclusters.  No such concentration is seen along other
coordinates although we see several peaks in the distribution of
clusters in both $Z$- and $Y$- directions.  We checked this
quantitatively by calculating the distribution of member clusters of
very rich superclusters along supergalactic coordinates
(Figure~6). In this way we can see whether the clusters are
concentrated in a certain supergalactic interval (this approach was
chosen because of simplicity and also because several rich systems
of superclusters are located almost parallel to one or another
plane of supergalactic coordinate axes).  The presence of the zone
of avoidance causes the absence of clusters around $Y = 0$,
therefore we can only compare the distributions of clusters along
the $X$ and $Z$ coordinates. In the case of uniform distribution
the distribution of clusters and superclusters along $X$ and $Z$
axes should be statistically identical.  However, the
Kolmogorov--Smirnov test shows that the zero hypothesis
(distribution of clusters  along $X$ and $Z$ coordinates is
identical) is rejected  at the 99\% confidence level.

We compared the distribution of the members of very rich
superclusters from real and random catalogues.  The results show
that of the 320 member clusters of observed very rich
superclusters 198 belong to the sheet $-75 \leq X \leq
25$~\Mpc. In the case of randomly located superclusters  the
expected number of clusters in very rich superclusters in the sheet
is 80 if we do not take into account the selection effects, and
123, if the selection effects have been taken into account.
Therefore no such concentration of clusters is seen in the case of
randomly located superclusters.

The evidence that the structures may be connected through the zone
of avoidance leads us to believe that superclusters in this
supergalactic $X$ interval form a {\it Dominant Supercluster
Plane}.  The figures show that this plane is almost perpendicular to
the $X$-axis and crosses the Supergalactic plane almost at right
angle.

In fact, already Tully \etal (1992) noted the presence of the
supercluster structures that are almost orthogonal to the
Supergalactic plane. Due to that they described the
supercluster-void network as a {\it three-dimensional chessboard}.
Our data show that structures delineated by rich superclusters are
not only orthogonal but also located quite regularly (Sections 6
and 7).  Thus, although the description as a chessboard is a
simplification it describes certain aspects of the
supercluster-void network rather well.

We list the superclusters belonging to the Dominant Supercluster
Plane: the Aquarius--Cetus, the Aquarius, the Aquarius B, the
Pisces--Cetus, the Horologium--Reticulum, the Sculptor, the
Fornax--Eridanus and the Caelum superclusters in the Southern sky,
and the Corona Borealis, the Bootes, the Hercules, the Virgo--Coma,
the Vela, the Leo, the Leo A, the Leo--Virgo and the Bootes A in
the Northern sky.  These superclusters do not form a featureless
wall -- the Dominant Supercluster Plane is formed by a number of
intertwined chains of rich superclusters.

\subsection{ The distribution of poor superclusters and isolated
clusters}

We showed that very rich superclusters are arranged in chains and
walls, separated by huge voids. To study the distribution of poor
superclusters and isolated clusters with respect to richer
superclusters we used the nearest neighbour test as in \ee. In this
test we calculate the distribution of distances of the nearest
neighbours between members of poor superclusters and isolated
clusters, and clusters belonging to rich superclusters, and the
distribution of distances between randomly located points and
clusters from rich superclusters.  In this way we can see whether these
clusters are located close to rich superclusters, or  they
form a more or less randomly distributed smooth population in
voids.

In order to obtain a hypothetical homogeneous void population we
generated a sample of random clusters which are located at a distance
$d > 24~h^{-1}$ Mpc from real clusters that belong to rich
superclusters and occupy the same volume as real clusters.
The number of these random
clusters was equal to the number of isolated clusters and poor
supercluster members. 

The results of this test are shown in the Figure~7. We see that the
nearest neighbour distribution curves of these sample pairs deviate
from each other -- real isolated clusters and members of poor
superclusters are located much closer
to rich superclusters than randomly located test clusters.
A Kolmogorov-Smirnov test shows that these distributions are
different at the 99 \% confidence level. In other words, isolated
clusters and clusters in poor superclusters belong to outlying
parts of superclusters and do not form a random population in
voids.

\begin{figure}[htbp] 
\epsfysize=8cm
\hbox{\epsfbox[90 90 400 500]{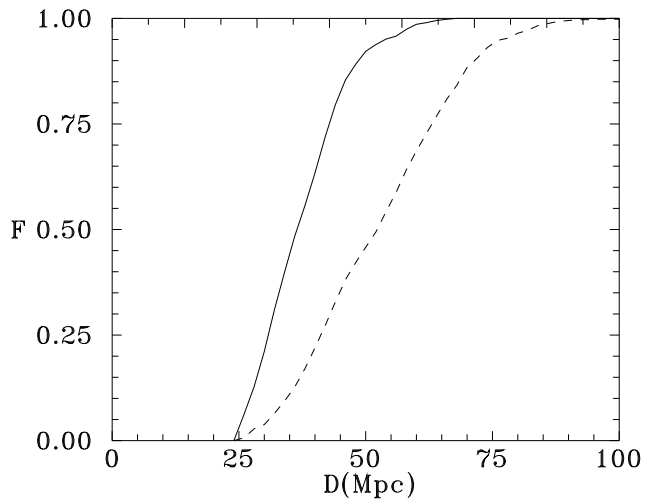}}
\caption {The integral probability distribution of the nearest
neighbour distances: cross distributions for the sample pairs of
clusters from rich 
superclusters vs. isolated clusters (solid curve) and
clusters from rich superclusters vs. random points (dashed curve). 
}
\end{figure}

\begin{table*}
\begin{center}
\caption{Median diameters of voids}
\begin{tabular}{rllllll}
\hline
$N_{cl}$&Sample&$D_{med}$&Sample&$D_{med}$&Sample&$D_{med}$ \\
        &      & \Mpc    &      & \Mpc    &      & \Mpc \\
\hline
1304 &  Acl  &  90 &          &           & Rcl   &$  88 \pm  6$\\
 900 &  Asc2& 100 &  Ard900&$ 94 \pm 5$ & Rsc2 &$ 144 \pm 26$\\
 580 &  Asc4& 110 &  Ard580&$102 \pm 9$ & Rsc4 &$ 173 \pm 20$\\
\hline
\end{tabular}
\end{center}
\end{table*}

\section{The sizes of voids between superclusters}

To calculate the sizes of voids between clusters and superclusters
we used the  {\it empty sphere method}. In this method we divide
the cubic sample volume into $n^3$ cubic cells, where $n$ is a
resolution parameter. For each cell centre we determine the
distance to the nearest cluster.  Cells having the largest
distances to the nearest clusters are located in centres of voids.
The distances to nearest clusters correspond to the radii of voids.
Therefore we obtain the void centre coordinates and radii. For
details of the method see Einasto, Einasto and Gramann (1989, EEG)
and \ee.

We determined the diameters of voids, delineated by all clusters, by
supercluster members, and by members of rich
superclusters (Table 1). The number of clusters in these samples
is 1304, 900, and 580 clusters, respectively, and samples are
denoted as Acl, Asc2, and Asc4 (A stands for Abell). In order to
see the influence of the change of the number of clusters on the
void sizes we used randomly diluted cluster samples, i.e. from the
observed sample (Acl) we removed clusters in a random way so that
in the resulting sample the number of clusters was 900 and 580
(correspondingly Ard900 and Ard 580, rd stands for random
dilution).

Additionally, we calculated void sizes for random supercluster
catalogues. Here again we used samples of all clusters, all
supercluster members and  members of rich superclusters 
(correspondingly the samples Rcl, Rsc2,
and Rsc4). We used ten realizations of random catalogues. Although
this number is rather small, the results for random catalogues are
seen quite well.  The median diameters of voids for these
catalogues are also given in the Table~1. Since the results of void
analysis for both sets of random supercluster catalogues were
essentially the same, we give in this Table the diameters of voids
for only one set, the censored random catalogues.

Table~1 shows that the median void sizes in the case of
observed cluster and supercluster samples are very close to each
other. Also the scatter of void diameter values is rather small
(see Fig.~8 in \ee).  We see only a slight increase of void sizes as
we move from the sample of all clusters to the supercluster members
and to the members of rich superclusters. The reason for the
increase of void sizes is clear: although isolated clusters and
poor superclusters are located close to void walls, some of them
enter into voids determined by rich superclusters, and thus voids
determined by all clusters are smaller -- the sizes of voids are
determined by the location of clusters in the
periphery of voids. If we remove clusters in a random way then of course
we remove part of the clusters from the central regions of void
walls that have no effect to void sizes.  Thus the increase of void
sizes in this case is smaller than in the first case. Real rich
superclusters  form a quasi-regular lattice which is almost
identical for supercluster samples of all richness classes; much
stronger random dilution is needed to destroy this lattice.

\begin{figure}[htbp]
\epsfysize=13cm
\hbox{\epsfbox[130 80 430 530]{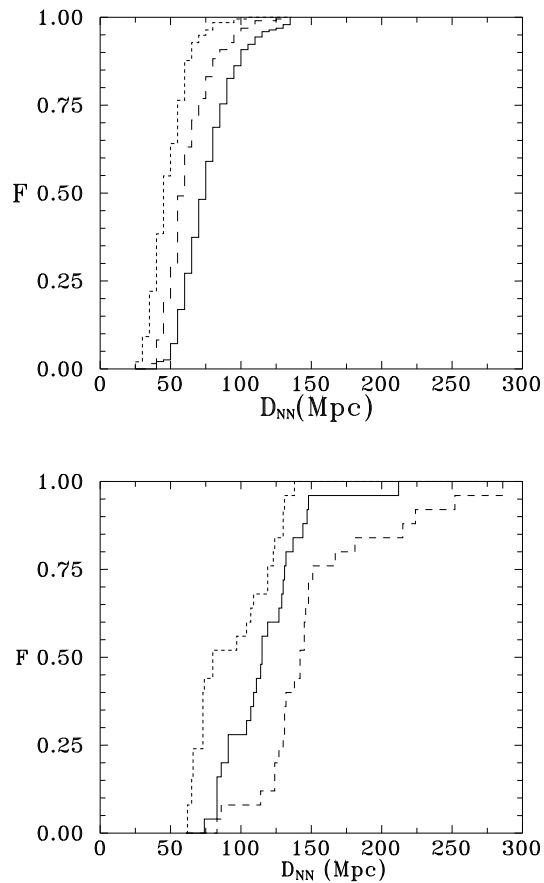}}
\caption {The distribution of distances between centres of
superclusters. Upper panel shows the distributions for poor and
medium rich superclusters, lower panel - for the very rich
superclusters.  Curves correspond to the first (line with short
dashes), second (line with long dashes) and third (solid line)
neighbour. }
\end{figure}

Comparison with the random catalogues shows that if the clusters
and superclusters are located randomly then the removal of part of
the clusters increases the  void sizes much more than in the
observed case.

\begin{table*}
\begin{center}
\caption{Median distances between high-density regions}
\begin{tabular}{rllllll}
\hline
$N_{cl}$& Sample &$D_{med}$& Sample &$D_{med}$& Sample &$D_{med}$ \\
        &        & \Mpc &           & \Mpc    &        & \Mpc    \\
\hline
1304 & Acl   & $122\pm 26$  &       &            & Rcl   & $ 143\pm 17 $\\
 900 & Asc2 & $126\pm 27$  & Ard900 & $120\pm 15$ & Rsc2 & $[139\pm 35]$\\
 580 & Asc4 & $[116\pm 18]$ & Ard580 & $122\pm 18$ &Rsc4 & $[140\pm 33]$\\
\hline
\end{tabular}
\end{center}
\end{table*}

\section{The characteristic distance between superclusters}

We saw that superclusters form intertwined systems that are
separated by giant voids of almost equal size. The characteristic
scale of this network can be calculated as a distance between
centres of superclusters on opposite sides of void walls.

We shall determine distances between high-density regions across
the voids using the {\it pencil-beam} analysis of mean distances
between high-density regions, as described by EEG.  The volume
under study was divided in one direction into $n^2$ beams. This
procedure was repeated in all three directions of coordinate axes,
therefore the total number of beams was $3 n^2$. In each beam we
used cluster analysis to determine density maxima and derived the
mean distance between two consecutive density maxima.  As a result
we obtained the mean distance between superclusters -- the mean
value over voids in all beams.  We used the neighbourhood radius $r
= 24~h^{-1}$, and two resolutions: $n = 24$ and $n = 12$.  The
lower resolution was used in the case of subsamples with smaller
number of clusters as will be described below.  To eliminate the
influence of the zone of avoidance we performed calculations
separately for the Northern and Southern sky.  This method finds
mean distances between systems independently of the supercluster
definition given in Section 3.  If the number of systems becomes
smaller then also the number of pencil-beams with systems detected
in them  becomes smaller. In that case we performed pencil-beam
analysis with a lower resolution, these results are given in
parenthesis.

Table~2 shows the results of our calculations. Mean distances between
high-density regions are given for the observed samples Acl, Asc2,
and Asc4, as well as for diluted samples Ard900 and Ard580, and
for random supercluster samples (first set of random samples) 
Rcl, Rsc2, and Rsc4 (the number of very rich superclusters in the
volume under study is too small to determine distances between them
using the pencil-beam method).  The Table shows that the distances
between systems for observed samples almost do not change. This is
understandable: in pencil-beams method we determine the positions
of the density maxima, and the presence of clusters in low-density
regions does not influence the results of this analysis. Thus the mean
separation of high-density regions across voids is almost identical
for all observed samples.  The same occurs in the case of randomly
located superclusters, only in this case distances between 
high-density regions are larger, and the number of detected
systems is about three times smaller than in the real case (most
beams cross none or only one high-density region and no distance
can be derived).

We can compare the last result with the direct estimate of 
the characteristic distance between high-density regions 
using void diameters determined above.  The median diameter of voids
delineated by members of superclusters was about $100~h^{-1}$ Mpc. If we
add the mean size of the shortest axis of the superclusters,
$20~h^{-1}$ Mpc (EETDA, Jaaniste \etal 1996) then we have as a
distance between supercluster centres across the voids a value of
120~\Mpc, close to that found using pencil-beam analysis.

We see that several tests indicate the presence of a characteristic scale of
about $120~h^{-1}$ Mpc in the distribution of rich clusters and superclusters
of galaxies. This scale corresponds to the distance between superclusters
across the voids. The small scatter of this distance enables us to say that
the supercluster-void network is rather regular. The present paper
confirms the results by EETDA based on a smaller dataset. This characteristic
scale is much larger than the typical scale of voids determined by galaxies
(Lindner \etal 1995), and is a manifestation of the hierarchy of
the distribution of galaxies and voids. Our data suggest also that
there exists no larger preferred scale in the Universe (cf. also
\ee). Thus the scale determined by the network of superclusters and
voids should be the upper end of the hierarchy of the distribution
of galaxies.

We shall discuss theoretical consequences of the presence of such a scale in
further papers of this series (Einasto \etal 1996b, Frisch \etal 1996).

{\section{Distribution of superclusters in void walls}}

Previous analysis has shown that the sizes of voids determined by
members of superclusters of different richness are almost
identical. This result, and the absence of a randomly located
population of clusters in voids, suggest that practically all
clusters are located in void walls, and the overall distribution of
superclusters of different richness is rather similar. Now we shall
study the distribution of superclusters of different richness in
void walls.  For that we calculate for each supercluster centre the
distances to the centres of three nearest superclusters, separately
for poor and medium rich, and for very rich superclusters
(Figure~8).

On the upper panel of Figure~8 these distributions are given for
poor and medium rich superclusters. We see, first, that these
distances are small, and second, that these distributions are
smooth and do not show the presence of any preferred distance
between superclusters (that would be seen as a peak in the distance
distribution). The median distances to the first, second and third
neighbours are, correspondingly, $D_{NN1} = 45$, $D_{NN2} = 60$ and
$D_{NN3} = 75$ \Mpc.

In the case of poor and medium rich superclusters from random
catalogues the median distances to the first, 
second and third neighbours are closer to each other than in the
observed case: $D_{NN1} = 60 \pm 3$,
$D_{NN2} = 63 \pm 3$ and $D_{NN3} = 70 \pm 1$ \Mpc.

The distributions of distances between very rich superclusters 
(Figure~8, lower panel) are quite different. None of these
distributions is smooth as in the upper panel. The most important
feature in this Figure is the presence of a peak in the distribution of
distances of the second and third neighbour in the interval $110 <
D_{NN2,3} < 150$ \Mpc\ -- over 75\% of very rich superclusters
have a neighbour at this distance. The median distances to the
second and third neighbours are, correspondingly, $D_{NN2} = 115$
and $D_{NN3} = 142$ \Mpc. These values are close to the size of
voids between superclusters.  

Since the number of very rich superclusters is small, it is easy to
check to which supercluster pairs these distances correspond. This
analysis confirms that the peak in the distribution of the second
and third neighbour is due to the pairs of superclusters on
opposite sides of void walls.

Also, this analysis shows that about half of the very rich
superclusters have their first nearest very rich neighbour at the
same side of the void (examples of such pairs are the
Fornax-Eridanus and the Caelum superclusters, the superclusters in
the Aquarius complex and others) at a distance less than $D_{NN1} =
75$ \Mpc.

One can argue that the last result may simply be due to the small
number of very rich superclusters. Thus we performed the same
analysis with superclusters from random catalogues. In this case
the distributions of neighbour distances for superclusters of all
richnesses are smoothly increasing without any strong peak as in the
observed case for very rich superclusters. The median distances
between centres of randomly located very rich superclusters are:
$D_{NN1} = 94 \pm 19$\Mpc, $D_{NN2} = 138 \pm 12$\Mpc, and
$D_{NN3} = 180 \pm 16$\Mpc,  values that are much larger than in
the observed case.

This test shows that the overall distribution of superclusters of
various richnesses is rather similar, but the distribution of
superclusters in void walls depends on the supercluster richness.

Additional evidence for differences in the distribution of poor 
and rich superclusters comes from the analysis of the correlation
function of clusters of galaxies (Einasto \etal 1996b).

\vskip 5mm

\section{Conclusions}

The main results of our analysis of the spatial distribution of
rich clusters and superclusters are:

\begin{itemize}
\item we present a new whole-sky catalogue of
superclusters of Abell-ACO clusters up to distances $z = 0.12$
which contains 220 superclusters, 90 of which have been determined
for the first time. There are several new very rich
superclusters with eight or more member clusters;

\item about 2/3 of very rich superclusters are located
in the Dominant Supercluster plane that is orthogonal to the
Supergalactic plane;

\item several tests suggest the presence of a characteristic
scale of about $120~h^{-1}$ Mpc  in the distribution of clusters
and superclusters of galaxies;

\item rich superclusters reside in chains and walls;

\item the distribution of superclusters in
void walls depends on the supercluster richness;

\item the mean space density of Abell-ACO clusters of galaxies,
corrected for incompleteness and Galactic extinction, is $2.6
\times$ 10$^{-5}$ $h^3$ {\rm Mpc}$^{-3}$.

\end{itemize}

\acknowledgements
This work was partly supported by the Estonian Science Foundation
(grant 182) and by the International Science Foundation (grant LDC
000). JJ would like to thank Division of Astronomy and Physics of
Estonian Academy of Sciences for financial support. We thank Drs
V. M\"uller, E. Saar and D. Tucker for fruitful discussions.

\end{document}